\documentclass[12pt,thmsa]{article}
\usepackage{amsfonts}
\usepackage{graphicx}

\input{tcilatex}
\begin{document}

\setcounter{page}{0} \topmargin 0pt \oddsidemargin 5mm \renewcommand{%
\thefootnote}{\fnsymbol{footnote}} \newpage \setcounter{page}{0} 
\begin{titlepage}
\begin{flushright}
Berlin Sfb288 Preprint  \\
hep-th/0002185\\
\end{flushright}
\vspace{0.5cm}
\begin{center}
{\Large {\bf Large and small Density Approximations to the thermodynamic 
Bethe Ansatz} }

\vspace{0.8cm}
{\large A. Fring and  C. Korff  }

\vspace{0.5cm}
{\em Institut f\"ur Theoretische Physik,
Freie Universit\"at Berlin\\ 
Arnimallee 14, D-14195 Berlin, Germany }
\end{center}
\vspace{0.2cm}
 
\renewcommand{\thefootnote}{\arabic{footnote}}
\setcounter{footnote}{0}

\begin{abstract}
We provide  analytical solutions to the thermodynamic Bethe ansatz
equations in the large and small density approximations. We extend results
previously obtained for leading order behaviour of the scaling function of 
affine Toda field theories related to simply laced Lie
algebras to the non-simply laced case. The comparison with semi-classical
methods shows perfect agreement for the simply laced case.
We derive the  Y-systems for affine Toda field theories with real coupling
constant
and employ them to improve the large density approximations. We test the quality 
of our analysis explicitly for the Sinh-Gordon model and the
$(G_2^{(1)},D_4^{(3)})$-affine Toda field theory. 
\end{abstract}
\vfill{ \hspace*{-9mm}
\begin{tabular}{l}
\rule{6 cm}{0.05 mm}\\
Fring@physik.fu-berlin.de \\
Korff@physik.fu-berlin.de \\
\end{tabular}}
\end{titlepage}
\newpage 


\section{Introduction}

In the context of 1+1-dimensional integrable quantum field theories numerous
methods have been developed to compute various quantities in an exact
manner, that is non-perturbative in the coupling constant. Sometimes it is
even possible to perform the related computations analytically. Often the
evaluation within one particular approach lacks information, typically a
constant, which might be supplied by an entirely different method. Ideally
one would like to achieve a situation in which each approach is
self-consistent.

An example for the situation just outlined is given for instance in the
form-factor program \cite{Kar}, which allows in principle to compute
correlation functions. The lowest non-vanishing form-factor is not fixed
within this approach and is typically obtained from elsewhere. For instance
the vacuum expectation value of the energy-momentum tensor can be extracted
from the thermodynamic Bethe ansatz (TBA) \cite{TBAZam1}. Alternatively, one
can compute correlation functions by perturbing around the conformal field
theory \cite{Zamocorr}. Also in this approach one appeals to the
thermodynamic Bethe ansatz for the vacuum expectation value of the
energy-momentum tensor and to the Bethe ansatz \cite{mum} for the relation
between the coupling constant and the masses. The latter correspondence is
also needed in an approach initiated recently in \cite{Zamref,Arsch,Fateev},
where it was observed that the ultraviolet asymptotic behaviour for many
theories may be well approximated by zero-mode dynamics. The considerations
in \cite{Zamref} exploit the knowledge of an exact reflection amplitude,
which on one hand results from certain manipulation on the three-point
function of the underlying ultraviolet conformal field theory and on the
other hand has a semi-classical counterpart in the related quantum
mechanical problem. The question why this approach allows to compute scaling
functions with a high accuracy is still to be settled \cite{Ztalk}.

The computation of scaling functions by means of the TBA does conceptually
not require any additional input from other methods. However, up-to-now it
is only possible to tackle the problem numerically due to the nonlinear
nature of the central equation involved. Several attempts have been made to
formulate analytical approximations. This is desirable for various reasons,
one being that the numerical effort becomes quite considerable for some
models with increasing particle species content. A further reason is of
course that analytical expressions allow to study further the deeper
structures of the theories. For instance for affine Toda field theories
(ATFT) \cite{ATFT} related to simply laced Lie algebras some analytical
expressions have been provided \cite{ZamoR,MM,FKS1}. In the approximation
method of \cite{ZamoR,MM,FKS1} a constant was left undetermined, which can
be fixed in the same spirit as outlined for the other methods in the
preceding paragraph, namely by appealing to another approach. Contrary to
the claim in \cite{Arsch}, we demonstrate in the present manuscript that it
is possible to fix the constant in this way without approximating higher
order terms. Extending the analysis of \cite{MM,FKS1} also to the non-simply
laced case in the present manuscript, we will demonstrate in addition that
the constant may also be well approximated from within the TBA-analysis.
Furthermore, we give simple analytical expressions for improved
approximations in the large and small density regime.

Our manuscript is organized as follows: In section 2 we provide large and
small density approximations for the solutions of the TBA-equations. In
section 3 we assemble the necessary data for ATFT needed to extend our
previous analysis to the non-simply laced case. We derive universal
TBA-equations and Y-systems for \emph{all} ATFT and show how they may be
utilized to improve on the analytical approximations. We derive the related
scaling function. We test the quality of the various approximations for the
explicit example of the Sinh-Gordon model and the $(G_{2}^{(1)},D_{4}^{(3)})$%
-ATFT. We state our conclusions in section 4.

\section{Large and small density approximations}

\subsection{The TBA}

The object of investigation of the TBA is a multiparticle system containing $%
n$ different particle species with masses $m_{i}$ ($1\leq i\leq n$) whose
dynamical interaction is described by a factorizable diagonal scattering
matrix $S_{ij}(\theta )$, which is a function of the rapidity difference $%
\theta $. We assume the statistical interaction to be of fermionic type%
\footnote{%
In order to keep the discussion as simple as possible we do not treat
general statistics here as for instance Haldane type \cite{BF}.
Generalizations of our arguments in this sense are straightforward.}.
Adopting the notation of \cite{FKS1} the thermodynamic Bethe Ansatz
equations \cite{TBAZam1}, which characterize the thermodynamic equilibrium
of such a system, are the $n$ coupled nonlinear integral equations 
\begin{equation}
rm_{i}\cosh \theta +\ln \left( e^{L_{i}(\theta )}-1\right)
=\sum\limits_{j=1}^{n}\left( \varphi _{ij}*L_{j}\right) (\theta )\,.
\label{TBAE}
\end{equation}
Here the scaling parameter $r$ is given by the inverse temperature times a
mass scale. The convolution of two functions is abbreviated as usual by $%
(f*g)(\theta ):=1/2\pi \int d\theta ^{\prime }\,f(\theta -\theta ^{\prime
})g(\theta ^{\prime })$. The TBA kernel reads 
\begin{equation}
\varphi _{ij}(\theta ):=-i\dfrac{d}{d\theta }\ln S_{ij}(\theta ).
\label{TBAkernel}
\end{equation}
The functions $L_{i}$, which are to be determined as solutions of the TBA
equations (\ref{TBAE}), are related to the particle densities $\rho _{r}^{i}$
and the densities of available states $\rho _{h}^{i}$ as $L_{i}=\ln (1+\rho
_{r}^{i}/\rho _{h}^{i})$, such that for physical reasons $L_{i}\geq 0$.
Keeping this definition in mind, we speak of the large density regime when $%
L_{i}>\ln 2$ and of the small density regime when $L_{i}<\ln 2$. It is
sometimes useful to express matters in terms the pseudo-energies $%
\varepsilon _{i}(\theta ):=-\ln [\exp (L_{i}(\theta ))-1)].$

Having solved the TBA-equations (\ref{TBAE}) for the $L$--functions one is
in principle in the position to evaluate the scaling function

\begin{equation}
c(r)=\dfrac{6r}{\pi ^{2}}\sum\limits_{i=1}^{n}m_{i}\dint_{0}^{\infty
}d\theta L_{i}(\theta )\cosh \theta  \label{scc}
\end{equation}
which can be interpreted as off-critical effective central charge belonging
to the conformal field theory obtained in the ultraviolet limit, i.e. $%
r\rightarrow 0$. It is our goal in this manuscript to approximate this
function in a simple analytical way to high accuracy.

\subsection{Approximative analytical Solutions}

In general it is possible to solve the TBA-equations numerically, where the
convergence of the iterative procedure is guaranteed by means of the Banach
fixed point theorem \cite{FKS1}. However, the numerical problem becomes
quite complex when one increases the number of particle species. For this
reason, and more important because one would like to gain a deeper
structural insight into the solutions of (\ref{TBAE}), it is desirable to
obtain analytical solutions to the TBA-equations. Due to the nonlinear
nature of (\ref{TBAE}) only few analytical solutions are known. Nonetheless,
one may obtain approximated analytical solutions when $r$ tends to zero. For
large $L_{i}$, i.e. for large particle densities, it was shown in \cite
{ZamoR,MM,FKS1} that the integral equation (\ref{TBAE}) may be turned into a
set of differential equations of infinite order. Under certain natural
assumptions, which are however not satisfied universally for all models, one
may approximate these equations by second order differential equations,
whose solutions are given by 
\begin{equation}
L_{i}^{0}(\theta )=\ln \left( \frac{\cos ^{2}\left( \beta _{i}\theta \right) 
}{2\beta _{i}^{2}\eta _{i}}\right) \quad \quad \quad \text{for }\left|
\theta \right| \leq \frac{\arccos (\beta _{i}\sqrt{2\eta _{i}})}{\beta _{i}}.
\label{Lapp}
\end{equation}
The restriction on the range of the rapidity stems from the physical
requirement $L_{i}\geq 0$. The $n$ constants $\eta _{i}=\sum_{j}\eta
_{ij}^{(2)}$ are determined by a power series expansion of the TBA kernel 
\begin{equation}
\widetilde{\varphi }_{ij}(t):=\int\nolimits_{-\infty }^{\infty }d\theta
\;\varphi _{ij}(\theta )e^{it\theta }=2\pi \sum_{n=0}^{\infty }(-i)^{n}\eta
_{ij}^{(n)}t^{n}\,.  \label{series}
\end{equation}
The dependence on the scaling parameter $r$ enters through the quantity 
\begin{equation}
\beta _{i}=\frac{\pi }{2(\delta _{i}-\ln (r/2))}\,\,.  \label{ccc}
\end{equation}
Here the $\beta _{i},\delta _{i}$ are constants of integration. There is a
very crude lower bound we can put immediately on $\delta _{i}$. From the
fact that $L_{i}^{0}(0)\geq \ln 2$, we deduce $\delta _{i}>1/\pi /\sqrt{\eta
_{i}}+\ln (r/2)$. For particular models we will provide below a rigorous
argument which establishes that in fact they do not depend on the particle
type, such that we may replace $\beta _{i}\rightarrow \beta $ and $\delta
_{i}\rightarrow \delta $. We will also show that they can be fixed by
appealing to the semi-classical approach in \cite{Zamref,Fateev}. In
addition, we provide an argument which determines them approximately from
within the TBA analysis by matching the large and small density regimes.
Since the constant turns out to be model dependent, we will report on it in
detail below when we discuss concrete theories.

The restriction on the range for the rapidities in (\ref{Lapp}), for which
the large density approximation $L_{i}^{0}(\theta )$ ceases to be valid,
makes it desirable to develop also an approximation for small densities. For
extremely small densities we naturally expect that the solution will tend to
the one for a free theory. Solving (\ref{TBAE}) for vanishing kernel yields
the well-known solution 
\begin{equation}
L_{i}^{f}(\theta )=\ln \left( 1+e^{-rm_{i}\cosh \theta }\right) \,\,.
\label{LF}
\end{equation}
Ideally we would like to have expressions for both regions which match at
some distinct rapidity value, say $\theta _{i}^{m}$, to be specified below.
Since $L_{i}^{0}(\theta )$ and $L_{i}^{f}(\theta )$ become relatively poor
approximations in the transition region between large and small densities,
we seek for improved analytical expressions. This is easily achieved by
expanding (\ref{TBAE}) around the ``zero order'' small density
approximations. In this case we obtain the integral representation 
\begin{equation}
L_{i}^{s}(\theta )=\exp \left( -rm_{i}\cosh \theta
+\sum\nolimits_{j=1}^{n}(\varphi _{ij}*L_{j}^{f})(\theta )\right) \,\,.\,
\end{equation}
For vanishing $\varphi _{ij}$ we may check for consistency and observe that
the functions $L^{s}(\theta )$ become the first term of the expansion in (%
\ref{LF}). One could try to proceed similarly for the large density regime
and develop around $L_{i}^{0}$ instead of $L_{i}^{f}$. However, there is an
immediate problem resulting from the restriction on the range of rapidities
for the validity of $L_{i}^{0}$, which makes it problematic to compute the
convolution. We shall therefore proceed in a different manner for the large
density regime and employ Y-systems for this purpose. 

In many cases the TBA-equations may be expressed equivalently as a set of
functional relations referred to as $Y$-systems in the literature \cite
{TBAZamun}. Introducing the quantities $Y_{i}=\exp (-\varepsilon _{i})$, the
determining equations can always be cast into the general form 
\begin{equation}
Y_{i}(\theta +i\pi \mu )Y_{i}(\theta -i\pi \mu )=\exp (g_{i}(\theta ))
\label{Yg}
\end{equation}
with $\mu $ being some real number and $g_{i}(\theta )$ being a function
whose precise form depends on the particular model. We can formally solve
the equation by Fourier transformations 
\begin{equation}
Y_{i}(\theta )=\exp \left[ (g_{i}*\gamma _{i})(\theta )\right] \,,\,\qquad
\qquad \gamma _{i}(\theta )=[2\mu \cosh (\theta /2/\mu )]^{1/2}  \label{soYg}
\end{equation}
i.e. substituting (\ref{soYg}) into the l.h.s. of (\ref{Yg}) yields $\exp
(g_{i}(\theta ))$. Of course this identification is not completely
compelling and we could have chosen also a different combination of $Y$'s.
However, in order to be able to evaluate the $g_{i}(\theta )$ we require a
concrete functional input for the function $Y_{i}(\theta )$ in form of an
approximated function. Choosing here the large density approximation $L^{0}$
makes the choice for $g_{i}(\theta )$ with hindsight somewhat canonical,
since other combinations lead generally to non-physical answers.

We replace now inside the defining relation of $g_{i}(\theta )$ the $Y$'s by 
$Y_{i}\left( {\theta }\right) \rightarrow \exp (L_{i}^{0}(\theta ))-1$.
Analogously to the approximating approach in \cite{ZamoR,MM,FKS1}, we can
replace the convolution by an infinite series of differentials 
\begin{equation}
\varepsilon _{i}(\theta )=-(g_{i}*\gamma _{i})(\theta
)\,=-\sum\limits_{m=0}^{\infty }\nu _{i}^{(m)}\frac{d^{m}}{d\theta ^{m}}%
g_{i}(\theta )\,\,\,,  \label{FTEg}
\end{equation}
where the $\nu $'s are defined by the power series expansion 
\begin{equation}
\int\limits_{-\infty }^{\infty }d\theta \gamma _{i}(\theta )\,e^{it\theta
}=2\pi \sum\limits_{m=0}^{\infty }(-i)^{m}\nu _{i}^{(m)}t^{m}=\pi
\sum\limits_{m=0}^{\infty }\frac{E_{2m}}{(2m)!}(\pi \eta
_{i})^{2m}\,t^{2m}\,\,.  \label{F2g}
\end{equation}
The $E_{m}$ denote the Euler numbers, which enter through the expansion $%
1/\cosh x=\sum_{m=0}^{\infty }x^{2m}E_{2m}/(2m)!$. In accordance with the
assumptions of our previous approximations for the solutions of the
TBA-equations in the large density approximation, we can neglect all higher
order derivatives of the $L_{i}^{0}(\theta )$. Thus we only keep the zeroth
order in (\ref{FTEg}). From (\ref{F2g}) we read off the coefficient $\nu
_{i}^{(0)}=1/2$, such that we obtain a simply expression for an improved
large density approximation 
\begin{equation}
L_{i}^{l}(\theta )=\ln [1+Y_{i}^{l}(\theta )]=\ln [1+\exp (g_{i}(\theta
)/2)]\,.  \label{LL}
\end{equation}
In principle we could proceed similarly for the small density approximation
and replace now $Y_{i}\left( {\theta }\right) \rightarrow \exp
(L_{i}^{s}(\theta ))-1$ in the defining relations for the $g_{i}$'s.
However, in this situation we can not neglect the higher order derivatives
of the $L_{i}^{s}$ such that we have to keep the convolution in (\ref{FTEg})
and end up with an integral representation instead. We now wish to match $%
L_{i}^{s}$ and $L_{i}^{l}$ in the transition region between the small and
large density approximations at some distinct value of the rapidity, say $%
\theta _{i}^{m}$. We select this point to be the value when the function $%
f_{i}(\theta )=(6/\pi ^{2})rm_{i}L_{i}(\theta )\cosh \theta $, which is
proportional to the free energy density for a particular particle species,
has its maximum in the small density approximation 
\begin{equation}
\left. \frac{d}{d\theta }f_{i}^{s}(\theta )\right| _{\theta _{i}^{m}}=0\,\,.
\label{detttt1}
\end{equation}
In regard to the quantity we wish to compute, the scaling function (\ref{scc}%
), this is the point in which we would like to have the highest degree of
agreement between the exact and approximated solution, since this will
optimize the outcome for $c(r)$. Having specified the $\theta _{i}^{m}$, the
matching condition provides a simple rational to fix the constant $\delta
_{i}$%
\begin{equation}
L_{i}^{l}(\theta _{i}^{m})=L_{i}^{s}(\theta _{i}^{m})\qquad \quad
\Rightarrow \quad \delta _{i}^{m}\,\,\,.  \label{detttt}
\end{equation}

Clearly, in general we can not solve these equations analytically, but it is
a trivial numerical problem which is by no means comparable with the one of
solving (\ref{TBAE}). Needless to say that the outcome of (\ref{detttt}) is
not to be considered as exact, but as our examples below demonstrate it will
lead to rather good approximations. One of the reasons why this procedure is
successful is that $L_{i}^{s}(\theta _{i}^{m})$ is still very close to the
precise solution, despite the fact that is at its worst in comparison with
the remaining rapidity range.

Combining the improved large and small density approximation we have the
following approximated analytical $L$-functions for the entire range of the
rapidity 
\begin{equation}
L_{i}^{a}(\theta )=\QATOPD\{ . {L_{i}^{l}(\theta )\qquad \qquad \text{for }%
\left| \theta \right| \leq \theta _{i}^{m}\,}{L_{i}^{s}(\theta )\qquad
\qquad \text{for }\left| \theta \right| >\theta _{i}^{m}\,\,}\,\,,\qquad
\qquad \qquad \qquad 
\end{equation}
such that the scaling function becomes well approximated by 
\begin{equation}
c(r)\simeq \sum\limits_{i=1}^{n}\dint_{0}^{\infty }d\theta f_{i}^{a}(\theta
)\,\,\,.
\end{equation}

To develop matters further and report on the quality of $L^{0}$, $L^{f}$, $%
L^{s}$, $L^{l}$ we have to specify a particular theory at this point.

\section{Affine Toda field theory}

Affine Toda field theories \cite{ATFT} form a well studied class of
relativistic integrable quantum field theories in 1+1 space-time dimensions.
To each of these field theories a pair of affine Lie algebras $(X_{n}^{(1)}%
\frak{,}\hat{X}^{(\ell )})$ \cite{Kac} is associated whose structure allows
universal statements concerning its properties, like the S-matrix, the mass
spectrum, the fusing rules, etc. Here $\hat{X}^{(\ell )}$ denotes a twisted
affine Lie algebra w.r.t. a Dynkin diagram automorphism of order $\ell $.
Both algebras are chosen to be dual to each other, i.e. $\hat{X}^{(\ell )}$
is obtained from the non-twisted algebra $X_{n}^{(1)}$ of rank $n$ by
exchanging roots and co-roots. For $X_{n}^{(1)}$ simply-laced both algebras
coincide, i.e. $X_{n}^{(1)}\cong \hat{X}^{(\ell )},\,\ell =1$, which is
reflected in the quantum theory by a strong-weak self-duality in the
coupling constant. Moreover, the mass spectrum renormalises by an overall
factor and the poles of the S-matrix in the physical sheet do not depend on
the coupling constant. For non-simply laced Lie algebras these features
cease to be valid. The quantum masses are now coupling dependent and flow
between the classical masses associated with $X_{n}^{(1)}$ and $\hat{X}%
^{(\ell )}$ in the weak and strong coupling limit, respectively.
Consequently, the physical poles of the S-matrix shift depending on the
coupling and the strong-weak self-duality is broken.

\subsection{The universal S-matrix}

Remarkably, despite these structural differences the S-matrix of ATFT can be
cast into a universal form covering the simply-laced as well as the
non-simply laced case \cite{Oota,FKS2}. For our purposes the formulation in
form of an integral representation is most useful

\begin{eqnarray}
S_{ij}(\theta ) &=&\exp \int_{0}^{\infty }\frac{dt}{t}\,\phi _{ij}(t)\,\sinh 
\frac{t\theta }{i\pi }\,,  \label{Sint} \\
\phi _{ij}(t) &=&8\sinh (t\vartheta _{h})\sinh (t_{j}\vartheta
_{H}t)\,\,\left( [K]_{q(t)\bar{q}(t)}\right) _{ij\quad }^{-1}\,\,\,.
\end{eqnarray}

\noindent Denoting a q-deformed integer $n$ as common by $%
[n]_{q}=(q^{n}-q^{-n})/(q^{1}-q^{-1})$, we introduced here a ``doubly
q-deformed'' version of the Cartan matrix $K$ \cite{Oota,FR,FKS2} of the
non-twisted Lie algebra 
\begin{equation}
\lbrack K_{ij}]_{q\bar{q}}=(q\bar{q}^{\,t_{i}}+q^{-1}\bar{q}%
^{\,-t_{i}})\delta _{ij}-[I_{ij}]_{\bar{q}}\,\,\quad   \label{cartan}
\end{equation}
for the generic deformation parameters $q,\bar{q}$. The incidence matrix $%
I_{ij}=2\delta _{ij}-K_{ij}$ of the $X_{n}^{(1)}$ related Dynkin diagram is
symmetrized by the integers $t_{i}$, i.e. $I_{ij}t_{j}=I_{ji}t_{i}$. With $%
\alpha _{i}$ being a simple root we fix the length of the long roots to be 2
and choose the convention $t_{i}=\ell \alpha _{i}^{2}/2$. Inside the
integral representation (\ref{Sint}) we take 
\begin{equation}
q(t)=e^{t\vartheta _{h}},\quad \bar{q}(t)=e^{t\vartheta _{H}},\quad \qquad 
\text{with}\quad \vartheta _{h}:=\frac{2-B}{2h}\,,\quad \vartheta _{H}:=%
\frac{B}{2H}
\end{equation}
for the deformation parameters, where $0\leq B\leq 2$ is the effective
coupling constant. We further need the Coxeter numbers $h,\,\hat{h}$ and the
dual Coxeter numbers $h^{\vee },\,\hat{h}^{\vee }$ of $X_{n}^{(1)}\frak{\ }$%
and $\hat{X}^{(\ell )}$, respectively, as well as the $\ell $-th Coxeter
number $H=\ell \hat{h}$ of $\hat{X}^{(\ell )}$. Complete tables of these
quantities for individual algebras may be found in \cite{Kac}.

\noindent The incidence matrix satisfies the relation \cite{Oota,FKS2} 
\begin{equation}
\sum_{j=1}^{n}\,[\,I_{ij}]_{\bar{q}(i\pi )}\,m_{j}=2\cosh (\theta
_{h}+t_{i}\theta _{H})\,m_{i}\,\,\,,  \label{mass}
\end{equation}
which will turn out to be crucial for the arguments below. We introduced
here the imaginary angles $\theta _{h}=i\pi \vartheta _{h}$ and $\,\theta
_{H}=i\pi \vartheta _{H}$.

\subsection{The TBA-kernel}

From the universal integral representation (\ref{Sint}), we can now
immediately derive the Fourier transformed TBA-kernel (\ref{series}) for
ATFT. However, when taking the logarithmic derivative one has to be careful
about interchanging the derivative with the integral, since these two
operations do not commute. Comparison with the block representation of the
S-matrix \cite{FKS2} yields

\begin{equation}
\varphi _{ij}(\theta )=-\frac{1}{2\pi }\int_{-\infty }^{\infty }dt\,\phi
_{ij}(t)\,\exp \frac{t\theta }{i\pi },  \label{logder}
\end{equation}
such that the Fourier transformed universal TBA-kernel (\ref{series})
acquires the form 
\begin{equation}
\widetilde{\varphi }_{ij}(t)=-\pi \phi _{ij}(\pi t)=-8\pi \sinh t\pi
\vartheta _{h}\sinh t_{j}\pi \vartheta _{H}\,\,\left( [K]_{q(\pi t)\bar{q}%
(\pi t)}\right) _{ij}^{-1}\,.  \label{uni2}
\end{equation}
To be able to carry out the discussion of the previous section we require
the second order coefficient $\eta _{ij}^{(2)}$ in the power series
expansion (\ref{series}). From (\ref{uni2}) we read off directly

\begin{equation}
\eta _{ij}^{(2)}=\frac{\pi ^{2}}{h\,H}\,B(2-B)K_{ij}^{-1}\,t_{j}=\frac{\pi
^{2}}{h\,h^{\vee }}\,B(2-B)\,(\lambda _{i}\cdot \lambda _{j})\,\,\,\,.
\end{equation}
In the latter equality we used the fact that the inverse of the Cartan
matrix is related to the fundamental weights as $\lambda
_{i}=\sum_{j}K_{ij}^{-1}\,\,\alpha _{j}$, $t_{i}=\ell \alpha _{i}^{2}/2$ and 
$H=\ell \,\hat{h}=\ell \,h^{\vee }$. This implies on the other hand that 
\begin{equation}
\eta _{i}=\frac{\pi ^{2}}{h\,h^{\vee }}\,B(2-B)\,(\lambda _{i}\cdot \rho )
\label{etta}
\end{equation}
with $\rho =\sum_{i}\lambda _{i}$ being the Weyl vector. Therefore 
\begin{equation}
\eta =\sum\limits_{i=1}^{n}\eta _{i}=B(2-B)\frac{\pi ^{2}\rho ^{2}}{%
h\,h^{\vee }}\,=nB(2-B)\frac{\pi ^{2}(h+1)}{12h\,}\,\,.  \label{eta2}
\end{equation}
We used here the Freudenthal-de Vries strange formula $\rho ^{2}=h^{\vee
}/12\,\dim X_{n}^{(1)}$ (see e.g. \cite{GO}) and the fact that $\dim
X_{n}^{(1)}=n(h+1)$. Thus we have generalized the result of \cite{FKS1} to
the non-simply laced case. Notice that in terms of quantities belonging to
the non-twisted Lie algebra $X_{n}^{(1)}$ the formula (\ref{eta2}) is
identical for the simply laced and the non-simply laced case.

\subsection{Universal TBA equations and Y-systems}

In analogy to the discussion for simply-laced Lie algebras \cite{FKS1}, the
universal expression for the kernel (\ref{uni2}) can be exploited in order
to derive universal TBA-equations for \emph{all} ATFT, which may be
expressed equivalently as a set of functional relations referred to as $Y$%
-systems. Fourier transforming (\ref{TBAE}) in a suitable manner and
invoking the convolution theorem we can manipulate the TBA equations by
using the expression (\ref{uni2}). After Fourier transforming back we obtain 
\begin{equation}
\varepsilon _{i}+\sum\limits_{j=1}^{n}\Delta
_{ij}*L_{j}=\sum\limits_{j=1}^{n}\,\Gamma _{ij}*(\varepsilon _{j}+L_{j})\,.
\label{uniTBA}
\end{equation}
The universal TBA kernels $\Delta $ and $\Gamma $ are then given by \qquad 
\begin{eqnarray}
\gamma _{i}(\theta ) &=&\left( 2(\vartheta _{h}+t_{i}\vartheta _{H})\cosh 
\tfrac{\theta }{2(\vartheta _{h}+t_{i}\vartheta _{H})}\right) ^{-1},\qquad 
\\
\Gamma _{ij}(\theta ) &=&\sum\limits_{k=1}^{I_{ij}}\gamma _{i}(\theta
+i(2k-1-I_{ij})\theta _{H}), \\
\Delta _{ij}(\theta ) &=&[\gamma _{i}(\theta +(\theta _{h}-t_{i}\theta
_{H}))+\gamma _{i}(\theta -(\theta _{h}-t_{i}\theta _{H}))]\,\delta _{ij}\,.
\label{kernel1}
\end{eqnarray}
The key point here is that the entire mass dependence, which enters through
the on-shell energies $m_{i}\cosh \theta $, has dropped out completely from
the equations due to the identity (\ref{mass}). Noting further that 
\begin{equation}
\,[\,I_{ij}]_{\bar{q}(i\pi )}\,m_{j}\cosh \theta
\,=\sum_{k=1}^{I_{ij}}\,m_{j}\,\cosh \left[ \theta +(2k-1-I_{ij})\theta
_{H}\right] \,,
\end{equation}
we have assembled all ingredients to derive functional relations for the
quantities $Y_{i}=\exp (-\varepsilon _{i})$. For this purpose we may either
shift the TBA equations appropriately in the complex rapidity plane or use
again Fourier transformations, see \cite{FKS1} 
\begin{equation}
Y_{i}\left( {\theta +}\theta _{h}+\,t_{i}\theta _{H}\right) Y_{i}\left( {%
\theta }-\theta _{h}-\,t_{i}\theta _{H}\right) =\tfrac{\,\left[
1+Y_{i}\left( \theta +\theta _{h}-t_{i}\theta _{H}\right) \right] \left[
1+Y_{i}\left( \theta -\theta _{h}+t_{i}\theta _{H}\right) \right] }{%
\prod\limits_{j=1}^{n}\prod\limits_{k=1}^{I_{ij}}\,\left[
1+Y_{j}^{\,-1}\left( \theta +(2k-1-I_{ij})\theta _{H}\right) \right] }\,\,.
\label{Y}
\end{equation}
These equations are of the general form (\ref{Yg}) and specify concretely
the quantities $\mu $ and $g_{i}(\theta )$. We recover various particular
cases from (\ref{Y}). In case the associated Lie algebra is simply-laced, we
have $\theta _{h}+t_{i}\theta _{H}\rightarrow i\pi /h$,$\quad \theta
_{h}-t_{i}\theta _{H}\rightarrow i\pi /h(1-B)$ and $I_{ij}\rightarrow 0,1$,
such that we recover the relations derived in \cite{FKS1}. As stated therein
we obtain the system for minimal ATFT \cite{TBAZamun} by taking the limit $%
B\rightarrow i\infty $.

The concrete formula for the approximated solution of the Y-systems in the
large density regime, as defined in (\ref{LL}), reads 
\begin{equation}
Y_{i}^{l}(\theta )=\tfrac{\cos (2\theta \beta _{i})+\cos (2(\theta
_{h}-t_{i}\theta _{H})\beta _{i})}{4\eta _{i}\beta _{i}^{2}}%
\prod\limits_{j=1}^{n}\prod\limits_{m=1}^{I_{ij}}\left( 1-\tfrac{2\eta
_{j}\beta _{j}^{2}}{\cos ^{2}(\beta _{j}(\theta +(2m-1-I_{ij})\theta _{H})}%
\right) ^{\frac{1}{2}}\,\,.  \label{ih}
\end{equation}

Exploiting possible periodicities of the functional equations (\ref{Y}) they
may be utilized in the process of obtaining approximated analytical
solutions \cite{TBAZam}. As we demonstrated they can also be employed to
improve on approximated analytical solution in the large density regime. In
the following subsection we supply a further application and use them to put
constraints on the constant of integration $\delta _{i}$ in (\ref{ccc}).

\subsection{The constants of integration $\beta $ and $\delta $}

There are various constraints we can put on the constants $\beta _{i}$ and $%
\delta _{i}$ on general grounds, e.g. the lower bound already mentioned.
Having the numerical data at hand we can use them to approximate the
constant. In \cite{FKS1} this was done by matching $L^{0}$ with the
numerical data at $\theta =0$ and a simple analytical approximation was
provided $\delta ^{\text{num}}=\ln [B(2-B)2^{1+B(2-B)}]$. Of course the idea
is to become entirely independent of the numerical analysis. For this reason
the argument which led to (\ref{detttt}) was given.

When we consider a concrete theory like ATFT, we can exploit its particular
structure and put additional constraints on the constants from general
properties. For instance, when we restrict ourselves to the simply laced
case it is obvious to demand that the constants respect also the strong-weak
duality, i.e. $\beta _{i}(B)=\beta _{i}(2-B)$ and $\delta _{i}(B)=\delta
_{i}(2-B)$.

Finally we present a brief argument which establishes that the constants $%
\beta _{i}$ are in fact independent of the particle type $i$. We replace for
this purpose in the functional relations (\ref{Y}) the Y-functions by $%
Y_{i}^{h}(\theta ))$ and consider the equation at $\theta =0$, such that 
\begin{equation}
\frac{\cosh ^{2}[\pi \beta _{i}(\vartheta _{h}+t_{i}\vartheta _{H})]-2\beta
_{i}^{2}\eta _{i}}{\cosh ^{2}[\pi \beta _{i}(\vartheta _{h}+t_{i}\vartheta
_{H})]}=\prod\limits_{j=1}^{n}\prod\limits_{m=1}^{I_{ij}}\,\tfrac{(\cosh
^{2}[\pi \beta _{i}(2m-1-I_{ij})\vartheta _{H}]-2\beta _{i}^{2}\eta _{i})^{%
\frac{1}{2}}}{\cosh [\pi \beta _{i}(2m-1-I_{ij})\vartheta _{H}]}\,\,.
\label{yy}
\end{equation}
Keeping in mind that $\beta _{i}$ is a very small quantity in the
ultraviolet regime, we expand (\ref{yy}) up to second order in $\beta _{i}$,
which yields after cancellation 
\begin{equation}
4t_{i}\vartheta _{h}\vartheta _{H}=\frac{\alpha _{i}^{2}}{2}\frac{B(2-B)}{%
hh^{\vee }}=\sum_{j=1}K_{ij}\frac{\beta _{j}^{2}}{\beta _{i}^{2}}\frac{\eta
_{j}}{\pi }=\frac{B(2-B)}{hh^{\vee }}\sum_{j=1}K_{ij}\frac{\beta _{j}^{2}}{%
\beta _{i}^{2}}(\lambda _{j}\cdot \rho )\,.  \label{bb}
\end{equation}
We substituted here the expression (\ref{etta}) for the constants $\eta _{j}$
in the last equality. Using once more the relation $\lambda
_{i}=\sum_{j}K_{ij}^{-1}\alpha _{j}$, we can evaluate the inner product such
that (\ref{bb}) reduces to 
\begin{equation}
\alpha _{i}^{2}=\sum_{j=1}^{n}\sum_{k=1}^{n}K_{ij}\frac{\beta _{j}^{2}}{%
\beta _{i}^{2}}K_{jk}^{-1}\alpha _{k}^{2}\,\,.  \label{aa}
\end{equation}
Clearly this equation is satisfied if all the $\beta _{i}$ are identical.
From the uniqueness of the solution of the TBA-equations follows then
immediately that we can always take $\beta _{i}\rightarrow \beta $. Since
the uniqueness is only rigorously established \cite{FKS1} for some of the
cases we are treating here, it is reassuring that we can obtain the same
result also directly from (\ref{aa}). From the fact that the $\beta _{i}$
are real numbers and all entries of the inverse Cartan matrix are positive
follows that $\beta _{i}^{2}=\beta _{j}^{2}$ for all $i$ and $j$. The
ambiguity in the sign is irrelevant for the use in $L(\theta )$.

\subsection{The Scaling Functions}

In \cite{FKS1} it was proven that the leading order behaviour of the scaling
function is given by 
\begin{equation}
c(r)\simeq n-\,\dfrac{3\eta }{(\delta -\ln (r/2))^{2}}=n\mathbf{\,}\left( 1-%
\dfrac{\pi ^{2}B(2-B)(h+1)}{4h(\delta -\ln (r/2))^{2}}\right) \,\,.
\label{capp}
\end{equation}
$\quad $ \noindent \noindent From our arguments in section 3.2, which led to
the general expression for the constant $\eta $ in form of (\ref{eta2}),
follows that in fact this expression holds for \emph{all} affine Toda field
theories related to a dual pair of simple affine Lie algebras $(X_{n}^{(1)}%
\frak{,}\hat{X}^{(\ell )})$. However, strong-weak duality is only guaranteed
for $\ell =1$.

Restricting ourselves to the simply laced case, we can view the results of 
\cite{Zamref,Arsch,Fateev} obtained by means of a semi-classical treatment
for the scaling function as complementary to the one obtained from the
TBA-analysis and compare directly with the expression (\ref{capp}).
Translating the quantities in \cite{Zamref,Fateev} to our conventions, i.e. $%
R\rightarrow r$, $B\rightarrow B/2$, we observe that $c(r)$ becomes a power
series expansion in $\beta $. We also observe that the second order
coefficients precisely coincide in their general form. Comparing the
expressions, we may read off directly 
\begin{equation}
\delta ^{\text{semi}}=\ln \left( \frac{4\pi \Gamma \left( \frac{1}{h}\right)
\left( \frac{2}{B}-1\right) ^{\frac{B}{2}-1}}{k\Gamma \left( \frac{1}{h}-%
\frac{B}{2h}\right) \Gamma \left( 1+\frac{B}{2h}\right) }\right) -\gamma
_{E}\,\,  \label{const}
\end{equation}
for all ATFT related to simply laced Lie algebras\footnote{%
The expressions in \cite{Zamref} and \cite{Fateev} only coincide if in the
former case $m=1$ and in the latter $m=1/2$. In addition, we note a missing
bracket in equation (6.20) of \cite{Zamref}, which is needed for the
identification. Replace $C\rightarrow $ $-4QC$ therein.}. Here $\gamma _{E}$
denotes Euler's constant and $k=(\prod_{i=1}^{l}n_{i}^{n_{i}})^{\frac{1}{2h}}
$ is a constant which can be computed from the Kac labels $n_{i}$ of the
related Lie algebra. Contrary to the statement made in \cite{Arsch}, this
identification can be carried out effortlessly without the need of higher
order terms. Recalling the simple analytical expression $\delta ^{\text{num}}
$ of \cite{FKS1} we may now compare. Figure 1 demonstrates impressively that
this working hypothesis shows exactly the same qualitative behaviour as $%
\delta ^{\text{semi}}$ and also quantitatively the difference is remarkably
small.

To illustrate the quality of our approximate solutions to the TBA-equations,
we shall now work out some explicit examples.

\subsection{Explicit Examples}

To exhibit whether there are any qualitative differences between the simply
laced and non-simply laced case we consider the first examples of these
series.

\subsubsection{The Sinh-Gordon Model}

The Sinh-Gordon model is the easiest example in the simply laced series and
therefore ideally suited as testing ground. The Coxeter number is $h=2$ in
this case. An efficient way to approximate the L-functions to a very high
accuracy is 
\begin{equation}
L^{a}(\theta )=\QATOPD\{ . {\ln \left[ 1+\frac{\cos (2\beta \theta )+\cosh
(\pi \beta (1-B))}{4\eta \beta ^{2}}\right] \qquad \quad \quad \text{for }%
\left| \theta \right| \leq \theta ^{m}}{\exp \left[ -rm\cosh \theta
+(\varphi *L^{f})(\theta )\right] \qquad \quad \text{for }\left| \theta
\right| >\theta ^{m}}\,\,
\end{equation}
with 
\begin{equation}
\varphi (\theta )=\frac{4\sin (\pi B/2)\cosh \theta }{\cosh 2\theta -\cos
\pi B},\qquad \eta =\frac{\pi ^{2}B(2-B)}{8}.
\end{equation}
\includegraphics[width=10cm,height=16cm,angle=-90]{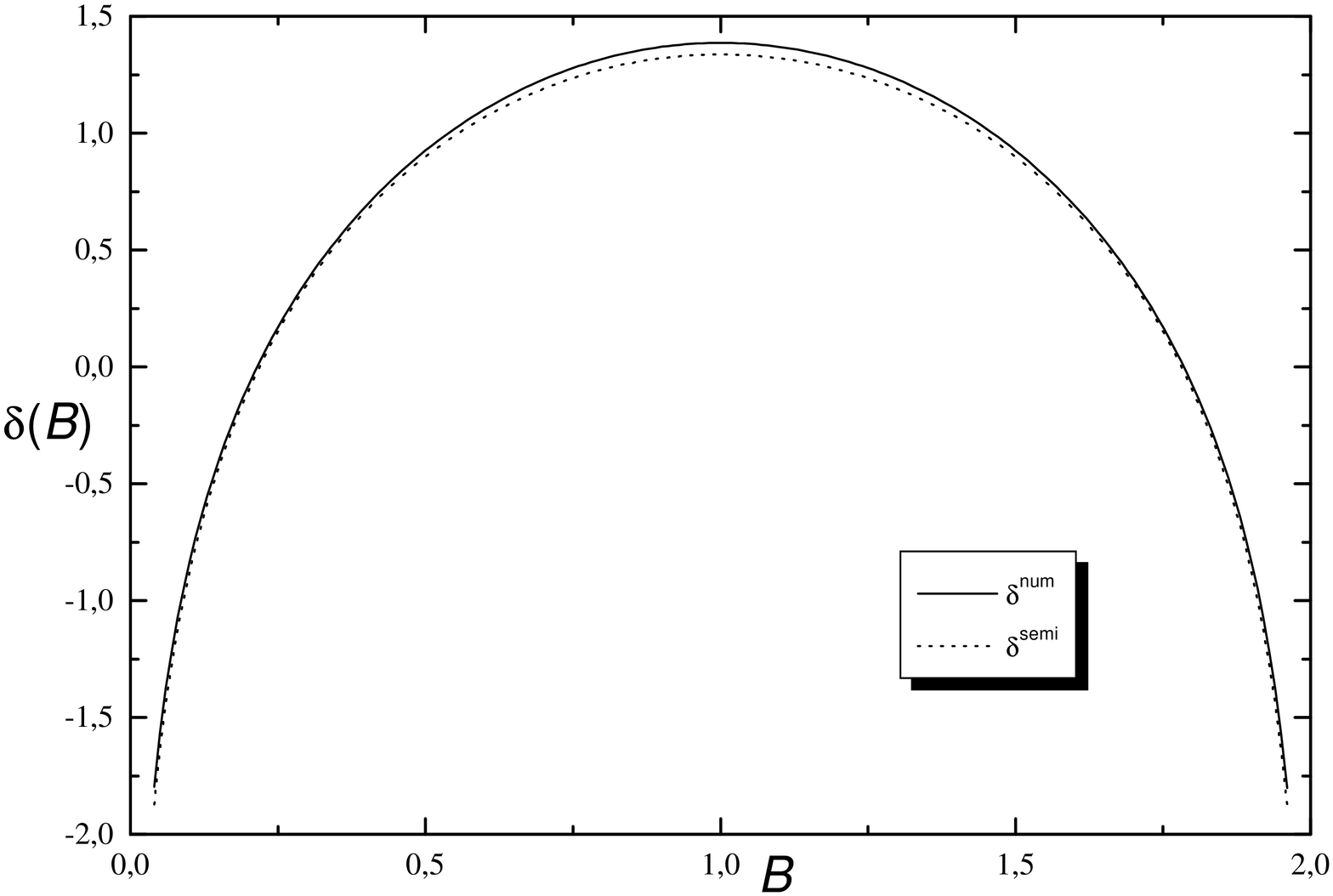}

\vspace*{0.7cm} 
\noindent {\small Figure 1: Numerically fitted constant }
$\delta ^{\text{num}}$ {\small versus the constant from the 
semi-classical approach for the Sinh-Gordon model 
$\delta ^{\text{semi}}$.}

The determining equation for the matching point reads 
\begin{equation}
\sinh \theta ^{m}-rm/2\sinh (2\theta ^{m})+\cosh (\theta ^{m})(\varphi
^{\prime }*L^{f})(\theta ^{m})=0\,\,\,.
\end{equation}
For instance for $B=0.4$ this equation yields $\theta ^{m}=11.9999$ such
that the matching condition (\ref{detttt}) gives $\delta ^{m}=0.4913$.
Figure 2(a) shows that the large and small density approximation $L^{0}$ and 
$L^{f}$ may be improved in a fairly easy way. In view of the simplicity of
the expression $L^{a}$ the agreement with the numerical solution is quite
remarkable. Figure 2(a) also illustrates that when using the constant $%
\delta ^{\text{semi}}$ instead of $\delta ^{m}$ the agreement with the
numerical solutions appears slightly better for small rapidities. When we
employ $\delta ^{\text{num}}$ instead of $\delta ^{\text{semi}}$ the
difference between the two approximated solutions is beyond resolution.
However, as may be deduced from Figure 2(b), with regard to the computation
of the scaling function the difference between using $\delta ^{m}$ instead
of $\delta ^{\text{semi}}$ is almost negligible. Whereas in the former case
the resulting value for the scaling function is slightly below the correct
value, it is slightly above by almost the same amount in the latter case.
More on the approximation of the scaling function in form of (\ref{capp})
may be found in \cite{FKS1}.

\subsubsection{$(G_{2}^{(1)},D_{4}^{(3)})$-ATFT}

In this case we have $h=6$ and $H=12$ for the related Coxeter numbers. The
two masses are $m_{1}=m\sin (\pi (1/6-B/24))$ and $m_{2}=m\sin (\pi
(1/3-B/12))$. The L-functions are well approximated by 
\[
L_{1}^{a}(\theta )=\QATOPD\{ . {\ln [1+\tfrac{\cos (2\beta \theta )+\cosh
(\pi \beta (\frac{1}{3}-\frac{B}{4}))}{4\eta _{1}\beta ^{2}}\sqrt{1-\tfrac{%
2\eta _{2}\beta ^{2}}{\cos ^{2}(\beta \theta )}}\,\,\,]\,\,\,\text{for }%
\left| \theta \right| \leq \theta _{1}^{m}}{\exp [-rm_{1}\cosh \theta
+(\varphi _{11}*L_{1}^{f}+\varphi _{12}*L_{2}^{f})(\theta )]\,\,\,\,\,\text{%
for }\left| \theta \right| >\theta _{1}^{m}}
\]
\[
\,\,L_{2}^{a}(\theta )=\QATOPD\{ . {\ln [1+\tfrac{\cos (2\beta \theta
)+\cosh (\pi \beta (\frac{1}{3}-\frac{5B}{24}))}{4\eta _{2}\beta ^{2}}%
\prod\limits_{k=-1}^{1}\sqrt{1-\tfrac{2\eta _{1}\beta ^{2}}{\cos ^{2}(\beta
(\theta +\frac{kB}{12}))}}\,]\,\text{for }\left| \theta \right| \leq \theta
_{2}^{m}}{\exp [-rm_{2}\cosh \theta +(\varphi _{21}*L_{1}^{f}+\varphi
_{22}*L_{2}^{f})(\theta )]\qquad \text{for }\left| \theta \right| >\theta
_{2}^{m}},
\]
with $\varphi $ given by (\ref{logder}) and 
\begin{equation}
\eta _{1}=\frac{5\pi ^{2}B(2-B)}{72},\quad \eta _{2}=\frac{\pi ^{2}B(2-B)}{8}%
,\quad \eta =\frac{7\pi ^{2}B(2-B)}{36}\,\,.
\end{equation}
Using now the numerical data $L_{1}(0)=4.2524$ and $L_{2}(0)=3.67144$ as
benchmarks, we compute by matching them with $L_{1}^{a}(0)$ and $L_{2}^{a}(0)
$ the constant to $\delta =1.1397$ in both cases. This confirms our general
result of section 3.4. Evaluating the equations (\ref{detttt}) and (\ref
{detttt1}) we obtain for $B=0.5$ the matching values for the rapidities $%
\theta _{1}^{m}=12.744$ and $\theta _{2}^{m}=12.278$ such that $\delta
_{1}^{m}=1.9539$ and $\delta _{2}^{m}=1.5572$. Figure 2(c) and 2(d) show a
good agreement with the numerical outcome.

The approximated analytical expression for the scaling function reads

\begin{equation}
c(r)\simeq 2-\dfrac{7\,\pi ^{2}B(2-B)}{12(\delta -\ln (r/2))^{2}}\,.
\end{equation}
This expressions differs from the one quoted in \cite{FKS1}, since in there
the sign of some scattering matrices at zero rapidity was chosen differently.

\includegraphics[width=10cm,height=16cm,angle=-90]{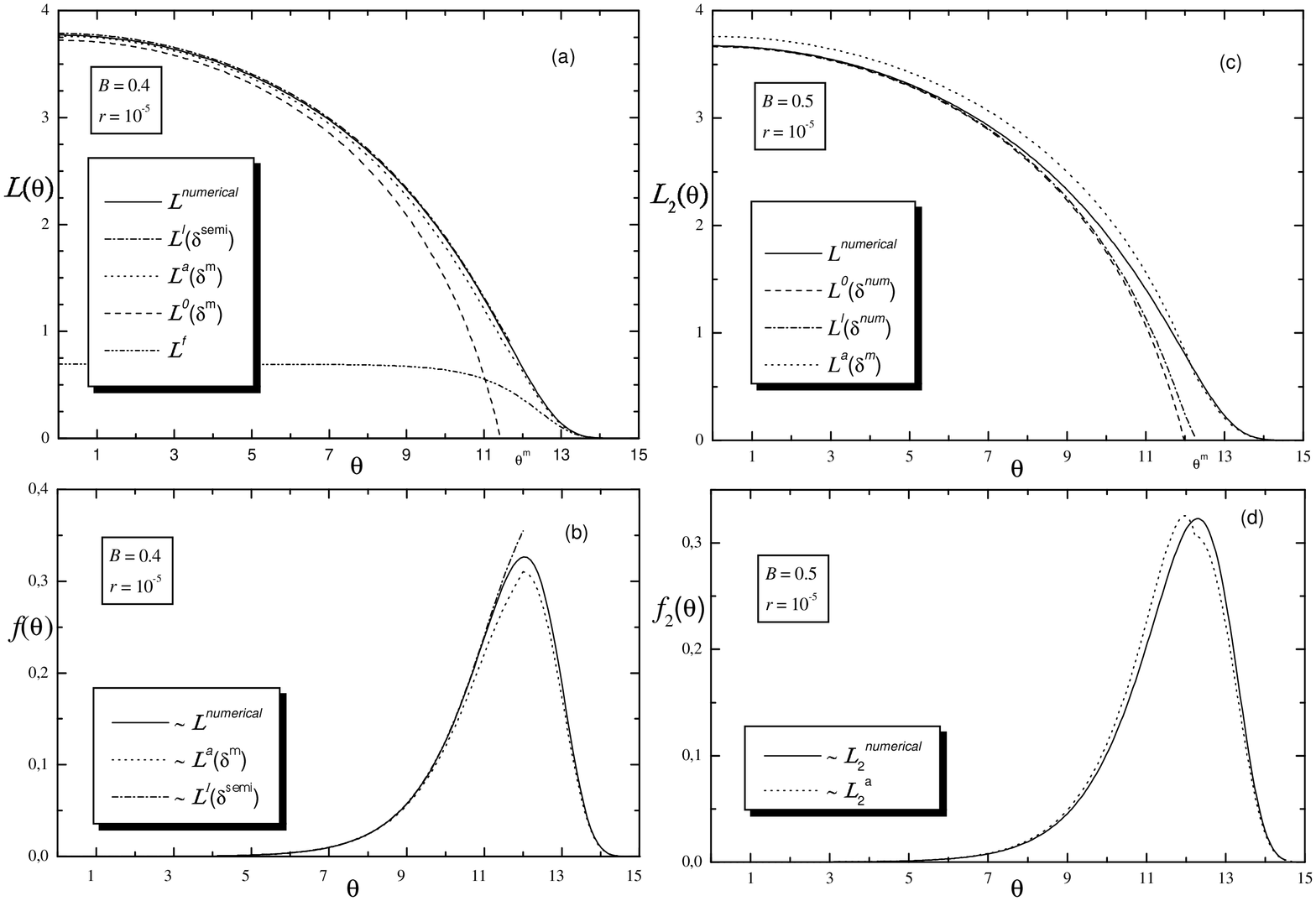}

\vspace*{0.6cm} 
\noindent {\small Figure 2: 
Various L-functions and free energy densities for the
Sinh-Gordon model (a), (b) at B=0.4 and r=10 $^{-5}$ and }$%
(G_{2}^{(1)},D_{4}^{(3)})${\small -ATFT at B=0.5 and r=10 $^{-5}$ (c), (d). }

\section{Conclusions}

We have demonstrated that it is possible to find simple analytical solutions
to the TBA-equation in the large and small density regime, which approximate
the exact solution to high accuracy. By matching the two solutions at the
point in which the particle density and the density of available states
coincide, it is possible to fix the constant of integration, which
originated in the approximation scheme of \cite{ZamoR,MM,FKS1} and was left
undetermined therein. Alternatively the constant may be fixed by a direct
comparison with a semi-classical treatment of the problem. It is not
necessary for this to proceed to higher order differential equations as was
claimed in \cite{Arsch}. Of course one may proceed further to higher orders,
but since the solutions to the higher order differential equations may only
be obtained approximately one does not gain any further structural insight
and moreover one has lost the virtue of the first order approximation, its
simplicity.

We derived the Y-systems for all ATFT and besides demonstrating how they can
be utilized to improve on the large density approximations we also showed
how they can be used to put constraints on the constant of integration. 

We have proven that the expression (\ref{capp}) for the scaling function is
of a general nature, i.e. valid for \emph{all} ATFT. It is desirable to
extend the semi-classical analysis \cite{Fateev} also to the non-simply
laced case. This would allow to read off  the constant $\delta $ also in
that case.

\vspace{0.5cm}

\textbf{Acknowledgments: } The authors are grateful to the Deutsche
Forschungsgemeinschaft (Sfb288) for financial support. We acknowledge
constructive conversations with B.J. Schulz at the early stage of this work
and are grateful to V.A. Fateev for kind comments.

\label{cc}

\begin{description}
\item  {\small \setlength{\baselineskip}{12pt}}
\end{description}

\end{document}